\documentclass[
 reprint,
 amsmath,amssymb,
 aps,onecolumn,
prstab,
]{revtex4-2}

\usepackage{graphicx}
\usepackage{dcolumn}
\usepackage{bm}
\usepackage{graphicx}

\usepackage{booktabs}
\usepackage{multirow}
\usepackage{threeparttable}
\begin{document}

\preprint{IMP/technote-202207- 01/Zhao}

\title{Study of the effects caused by space charge in electron cooling}

\author{He Zhao}
 \email{hezhao@impcas.ac.cn}
\affiliation{Institute of Modern Physics, Chinese Academy of Sciences, Lanzhou 730000, China}
\affiliation{University of Chinese Academy of Sciences, Beijing 100049, China}

\date{\today}


\begin{abstract} 
In electron cooling, the space charge (SC) is an important effect, which will affect the e-beam velocity distribution and thus the cooling performance. In this paper, we analyse several important effects that due to the space charge field, such like transverse and longitudianl space charge force, drift velocity caused by the SC and magnetic field, and longitudinal momentum deviation due to the SC during acceleration. 

\end{abstract}

\maketitle


\section{introduction}

Space charge is the most fundamental effect in the collective effects whose impact generally is proportional to the beam intensity. The charge and current of the beam create self-field (direct space charge) and image field (indirect space charge) which alter its dynamic behaviour and influence the single particle motion as well as coherent oscillations of the beam as a whole. In this note, we just talk about the direct space charge inside the beam. 

The space charge effect has been well studied and modeled years ago for the beam with several certain distributions, such like Uniform, Gaussian, Elliptical, Parabolic, etc \cite{handbook}. But all of them are point to the collective instability study on the intensity beam. According to the author's acknowledge, the self-field related effects haven't been well investigated in electron cooling \cite{e-cooling1,e-cooling2}, especially for the hollow electron beam \cite{hollow-e}. In electron cooling, the space charge force on e-beam and ion beam, the drift velocity and longitudinal velocity deviation of e-ebeam are all affected by the space charge effect, as well as the longitudinal magnetic field along the cooling section. We will focus on these studies in this note.


\section{Space charge induced effects in e-cooling }

\subsection{space charge force}

In an tranditional electron cooler, the e-beam is generally considered to be a round beam in transverse, and infinity long in longitudinal comparing to the ion beam. Assuming a round electron beam, the local beam density can be described by $\rho(r,s) = f(r)\lambda(s)$ with $f(r)$ the transverse distribution and $\lambda(s)$ the linear density in the longitudinal direction. The function $f(r)$ should satisfy the normalization condition,
\begin{equation}
	\int_{0}^{R_e} 2\pi r f(r) dr= 1,
\end{equation}
where $R_e$ is the radius of the electron beam. Here we think a DC e-beam with the current $I_e$, so we have
\begin{equation}
	I_e=\frac{Q}{t}= e \beta c \lambda \int_0^{R_e} 2 \pi r f(r) dr = e \beta c \lambda.
\end{equation}
Then we know the beam number density (DC beam) is
\begin{equation}
	\rho(r) =  \frac{I_e}{e \beta c } f(r).
\end{equation}

Based on the beam density, the electric and magnetic fields can be derived according to Maxwell equations. Firstly, Gauss's law gives us $\oint_s E_{sc}(r) ds=Q/\varepsilon_0$, so we get
\begin{equation}
	2 \pi r E_{sc}(r) \Delta L  = \frac{e  \int_0^{r} 2 \pi x \rho(x) dx} {\varepsilon} \Delta L,
\end{equation}
\begin{equation}
	E_{sc}(r)  = \frac{e \int_0^{r} x \rho(x) dx} {\varepsilon_0 r}.
\end{equation}
Also according to Ampère's circuital law $\oint_l B_{\phi}(r) dl = \mu_0 I(r)$ and $\mu_0 \varepsilon_0=1/ c^2$, we know
\begin{equation}
	2 \pi r B_{\phi}(r)  = \mu_0 e \beta c \int_0^{r} 2 \pi x \rho(x) dx,
\end{equation}
\begin{equation}
	B_{\phi}(r)  = \frac{e \beta \int_0^{r} x \rho(x) dx} {\varepsilon_0 r c}.
\end{equation}
The Lorenz force that the e-beam generated is given by
\begin{equation}
	F(r)=e Z_i (\mathbf{E_{sc}} + \mathbf{v} \times \mathbf{B_{\phi}})=\frac{e^2 Z_i \int_0^{r} x \rho(x) dx}{\varepsilon_0 \gamma^2 r}.
\end{equation}
It shows that the magnetic field partially cancels the electrostatic field, which complete cancellation at $v_s=c$. The space charge effect is significant at low beam energies.

In above, there is no longitudinal space charge because we use a DC e-beam ($\lambda(s)=constant$). Otherwise, the changing of e-beam current in longitudinal direction will also cause space change field. For a perfectly conducting beam pipe, it can be calculated by 
\begin{equation}
 E(s) = \frac{e g}{4\pi \varepsilon_0 \gamma^2} \frac{\partial \lambda(s)}{\partial s},
\end{equation}
where the geometric factor g depends on the beam distribution in transverse direction,
\begin{equation}\label{eq-g-factor}
 g = \int_r^b \frac{2}{x} \int_0^x 2 \pi r f(r)dr dx \simeq \int_0^b \frac{2}{x} \int_0^x 2 \pi r f(r)dr dx,,
\end{equation}
where $b$ is the radius of beam pipe. Acctually, the geometric factor is dependent on the radial position. Normally, since the ion beam size is quite smaller than the pipe, we only consider the electric field on axis. This factor for beams with various distributions have been well modeled in Ref. \cite{SC-Impedance}.

\subsection{Drift velocity}

In electron cooling, a longitudinal magnetic field is usually applied on the electrons, which is essential for the e-beam adiabatic expansion \cite{expansion}, e-beam focusing \cite{focusing} and magnetized cooling \cite{mag_cooling}, etc. As a consequence, a azimuthal drift velocity $v_{drift}$ will be generated by these two fields, i.e. $\bm{E}\times\bm{B}$ drift, which introduces the extra temperature of the e-beam. The formula of this drift velocity for electrons is

\begin{equation}
	v_{drift}(r) = \frac{\mathbf{E_{sc}} \times \mathbf{B_g}} {\mathbf{B_g^2}} = \frac{E_{sc}(r)}{B_g} \mathbf{\bm{\hat{r}} \times \bm{\hat{s}}} = \frac{e \int_0^{r} x \rho(x) dx} {\varepsilon_0 r B_g} \mathbf{\bm{\hat{r}} \times \bm{\hat{s}}}.
\end{equation}
where $\mathbf{E_{sc}}$ is the space charge field and $\mathbf{B_g}$ is the longitudinal magnetic field. Since the cooling force mainly depends on the velocity distribution of e-beam, it is necessary to include the drift velocity in the cooling calculation.

\subsection{Deviation of longitudinal velocity}

Because of the space charge depression, electrons at different radii inside the beam have different longitudinal velocities after acceleration. Accordingly, the relative longitudinal velocity component of an ion at a certain radius in the cooling calculation should be defined with respect to that of an electron at the same radius. Based on the radial electric field, the potential difference between the electron at radius r and the center is
\begin{equation}
	\Delta U = - \int_0^r E_{sc}(r) dr.
\end{equation}
We know that $\frac{\Delta E}{E}=\frac{e\Delta U}{\gamma E_0}$, therefore the velocity deviation of an electron at radius r from that on axis is
\begin{equation}
	\frac{\Delta v_s(r)}{v_{s0}} = \frac{1}{\gamma^2 \beta^2} \frac{\Delta E}{E} = \frac{e} {\gamma^3 \beta^2 E_0} \int_0^r E_{sc}(r) dr = \frac{e} {\gamma^3 \beta^2 E_0} \left \{ \int_0^r \left [ \frac{e}{\varepsilon_0 r} \int_0^{r} x \rho(x) dx \right ] dr \right \}
\end{equation}
This is another important factor that needs to be included in the cooling calculation.

\section{calculation}

According to above, we know the space charge force, drift velocity and longitudinal velocity deviation are all depended on the integral of the beam density. They can be descirbed by:
\begin{equation}
\begin{aligned}
	F(r) & \propto \frac{\int_0^{r} x \rho(x) dx}{ r} \\
	v_{drift}(r) & \propto \frac {\int_0^{r} x \rho(x) dx}{ r}\\
	\frac{\Delta v_s(r)}{v_{s0}} & \propto  \int_0^r \left [\frac{1}{r} \int_0^{r} x \rho(x) dx \right ] dr \\
\end{aligned}
\end{equation}
For several typical e-beam distributions such like uniform, ellitical, Gaussian and Hollow, it would be useful to give the analytical results. We did these derivation as summarized in Table \ref{tab-1}. We see that the analytical results are not always that simple, it would be much easier to use the numerical calculation for some special cases.  

\begin{table}[hbt]
\caption{\label{tab-1} Analytical results for several e-beam distributions}
\begin{threeparttable}
	\begin{tabular}{l c c c}
	\hline\hline
	    & Radial distribution $f(r)$  & $\int_0^{r} x f(x) dx/r ,\ (0<r<R_e)$ & $ \int_0^r [\int_0^{r} x f(x) dx/r] dr ,\ (0<r<R_e)$  \\
	\hline
	\specialrule{0pt}{4pt}{4pt}
	Uniform	& $\frac{1}{\pi R_e^2} \theta (R_e -r) \tnote{}$	& $\frac{r}{2\pi R_e^2}$ & $\frac{r^2}{4 \pi R_e^2}$  \\
	\specialrule{0pt}{6pt}{6pt}
	Elliptical  & $\frac{3}{2 \pi R_e^2} \left( 1 - \frac{r^2}{R_e^2} \right )^{1/2} \theta (R_e -r)$		& $\frac{1}{2\pi r} \left[ 1 - \left( 1- \frac{r^2}{R_e^2} \right)^{3/2} \right]$ & $ \frac{ \frac{4}{3} + \sqrt{1-\frac{r^2}{Re^2}} \left( \frac{r^2}{3Re^2} -\frac{4}{3} \right) +Ln \left[ \frac{1}{2} \left( 1+\sqrt{1-\frac{r^2}{R_e^2}}  \right) \right]}{2 \pi}$   \\
	\specialrule{0pt}{6pt}{6pt}
	Parabolic	& $\frac{2}{\pi R_e^2} \left( 1-\frac{r^2}{R_e^2} \right) \theta (R_e -r) $ & $\frac{r}{\pi R_e^2} \left( 1 - \frac{r^2}{2 R_e^2} \right)$ & $\frac{r^2}{2 \pi R_e^2} \left( 1 - \frac{r^2}{4 R_e^2} \right)$ \\
	\specialrule{0pt}{6pt}{6pt}
	Cos-square	& $\frac{2 \pi}{(\pi^2-4)R_e^2} Cos^2(\frac{\pi r}{2 R_e}) \ \theta (R_e -r) $ & $ \frac{\pi^2 r^2 - 2R_e^2 [1 - Cos(\frac{\pi r}{R_e})] + 2 \pi r R_e Sin(\frac{\pi r}{R_e})}{2 \pi (\pi^2 - 4) R_e^2 r} $ & $\frac{\pi^2 r^2 - 4(\gamma_e -1) R_e^2 + 4R_e^2 [ Ci(\frac{\pi r}{R_e}) -Cos(\frac{\pi r}{R_e}) - Ln(\frac{\pi r}{R_e})] }{4\pi (\pi^2-4) R_e^2} $ \\
	\specialrule{0pt}{6pt}{6pt}
	Gaussian	& $\frac{1}{2 \pi \sigma_r^2} e^{-\frac{r^2}{2 \sigma_r^2} }$ &  $\frac{1}{2 \pi r} \left( 1 - e^{-\frac{r^2}{2 \sigma_r^2} } \right) $  & $\frac{1}{4 \pi} \left[ \gamma_e + Ln(\frac{r^2}{2 \sigma_r^2}) + E_1(\frac{r^2}{2 \sigma_r^2})  \right] $ \\
	\specialrule{0pt}{6pt}{6pt}
	\multirow{2}*{Hollow}	& \multirow{2}*{$\frac{A_N}{2 \pi \sigma_r^2} e^{-\frac{(r-r_{0})^2}{2 \sigma_r^2}} \tnote{}$ } & $ \frac{A_N}{2 \pi r}  \left[ e^{-\frac{r_0^2}{2\sigma_r^2}} -e^{-\frac{(r-r_0)^2}{2\sigma_r^2}}\right]  $  &  \multirow{2}*{$Ln(r) + \sum\limits_{n=0}^{\infty} \frac{(r-r_0)^{2n+2} \ _{2}F_1 (1,2n+2;2n+3;1-\frac{r}{r_0})}{\sqrt{\pi} n! (-1)^n (2n^2+3n+1)(\sqrt{2}\sigma_r)^{2n+1}r_0} $}  \\
	\specialrule{0pt}{1pt}{1pt}
			&	& $ + \frac{A_N r_0}{2 \sqrt{2 \pi} \sigma_r r}  \left[ Erf(\frac{r_0}{\sqrt{2}\sigma_r}) +Erf(\frac{r-r_0}{\sqrt{2}\sigma_r})\right]  $  &  \\
	\specialrule{0pt}{2pt}{2pt}
	\hline\hline
	\end{tabular}

	\begin{tablenotes}
		\item[1]  Heaviside theta function $\theta(x)=1 \ for \ x>0, \ and \ \theta(x)=0 \ for \ x<0$
		\item[2]  $A_N = \sigma_r / \{ r_0\sqrt{\pi/2}[1+Erf(\frac{r_0}{\sqrt{2}\sigma_r})] + \sigma_r exp(-\frac{r_0^2}{2\sigma_r^2}) \}$
		\item[3] $Erf(x)$ is the error function.
		\item[4] $\gamma_e\approx 0.57721566$ is the Euler's constant.
		\item[5] The trigonometric (consine) integral $Ci(z)=-\int_z^{\infty} cos(t) /tdt$
		\item[6] The exponential integral $E_1(z)=\int_z^{\infty} e^{-t}/t dt$
		\item[7] The hypergeometric function $_2F_1(a,b;c;z) = \sum\limits_{k=0}^{\infty} \frac{a^k b^k}{c^k} \frac{z^k}{k!} $
 	\end{tablenotes}
\end{threeparttable}
\end{table}

\begin{figure}[hbt]
	\includegraphics[width=10cm]{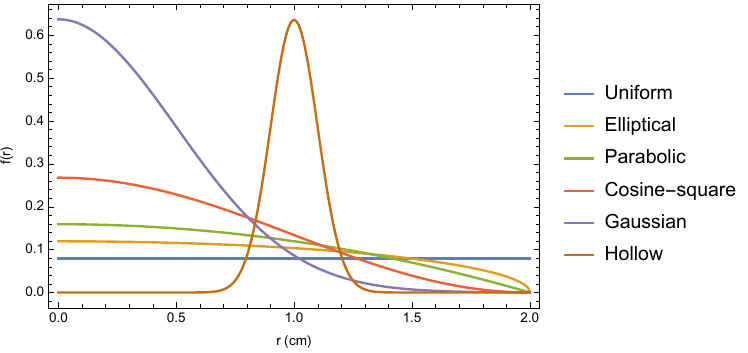}
	\includegraphics[width=10cm]{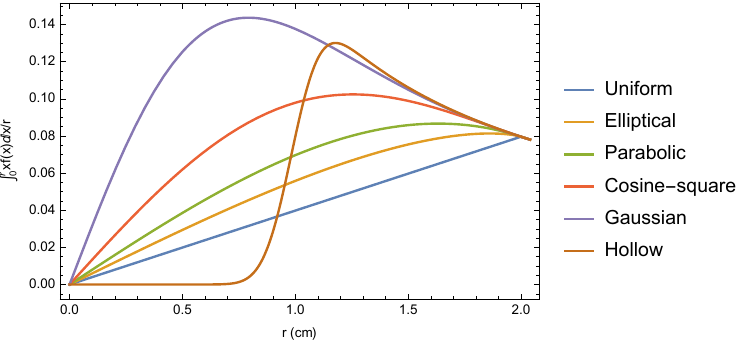}
	\includegraphics[width=10cm]{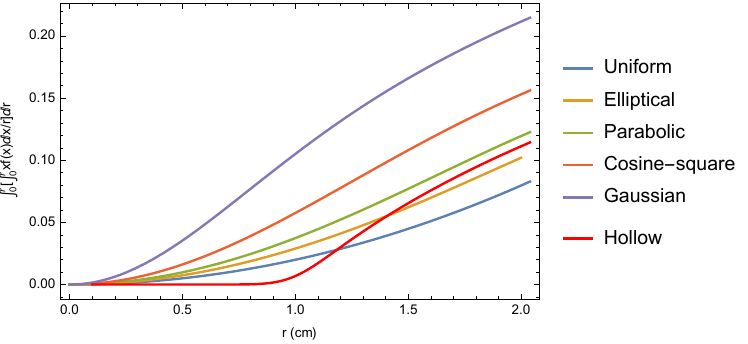}
	\caption{\label{fig:r_dis} Various beam transverse distribution $f(r)$ with beam radius $R_e=2\ cm$ with $\sigma_r$=0.5 cm for Gaussian beam, and $r_0$ = 1.0 cm and $\sigma_r$ = 0.1 cm for hollow beam.}
\end{figure}

According to the analytical results that list in Table~\ref{tab-1}, the integrals for various beam distributions are calculated as shown in Fig. (\ref{fig:r_dis}), in which the beam radius is $R_e=2\ cm$ with $\sigma_r$=0.5 cm for Gaussian beam, and $r_0$ = 1.0 cm and $\sigma_r$ = 0.1 cm for hollow beam. These results have been checked by numerical simulation.

\section{summary}
Several important effects that due to the space charge field in electron cooling are analysed, such like the transverse and longitudianl space charge force, drift velocity and longitudinal momentum deviation. These effects are important for the conventional electron cooling because the beam energy is quite low and the space charge field is strong enough to have a noticeable effect on the beam distribution.



\end{document}